\documentclass[aps,prl,twocolumn,showpacs]{revtex4}
\usepackage{graphicx}  
\usepackage{dcolumn}   
\usepackage{bm}        
\usepackage{amssymb}   
\usepackage{braket}

\newtheorem{theorem}{Observation}

\newcommand{\EE}{\ensuremath{\mathcal{E}}}
\renewcommand{\vr}{\ensuremath{\varrho}}
\newcommand{\tr}{\text{tr}}
\newcommand{\TR}{\textrm{tr}}

\begin{document}

\title{Joint measurability of generalized measurements implies classicality}
\date{\today}

\author{Roope Uola}
\author{Tobias Moroder}
\author{Otfried G\"uhne}
\affiliation{Naturwissenschaftlich-Technische Fakult\"at, Universit\"at Siegen, Walter-Flex-Str. 3, 57068
Siegen, Germany}

\begin{abstract}
The fact that not all measurements can be carried out simultaneously 
is a peculiar feature of quantum mechanics and responsible for many key 
phenomena in the theory, such as complementarity or uncertainty relations. 
For the special case of projective measurements quantum behavior can be 
characterized by the commutator but for generalized measurements it is not
easy to decide whether two measurements can still be understood in classical 
terms or whether they show already quantum features.  We prove that 
generalized measurements which do not fulfill the notion of joint 
measurability are nonclassical, as they can be used for the task of 
quantum steering. This shows that the notion of joint measurability is, among several definitions, 
the proper one to characterize quantum behavior. Moreover, the equivalence 
allows to derive novel steering inequalities from known results on joint 
measurability and new criteria for joint measurability from known results 
on the steerability 
of states.

\end{abstract}

\maketitle

{\it Introduction ---} Quantum theory is formulated in the language of Hilbert 
spaces, where states correspond to vectors or density matrices, and measurements 
are described by Hermitian matrices, the so-called observables. As realized 
by M. Born and P. Jordan, two observables $A$  and $B$ do not necessarily 
commute, which means in first place that the corresponding measurements cannot 
be carried out simultaneously in a direct way \cite{born, heisenberg}. This 
intuition can be made precise by formulating uncertainty relations, where 
the commutator $[A,B]=AB-BA$ quantifies the degree of uncertainty about the 
values of $A$ and $B$ \cite{heisenberg, kennard, robertson}. Consequently 
there is the widespread opinion that noncommuting observables are central 
for many quantum effects, while commuting observables are considered to be 
classical. 

It has turned out, however, that the notion of observables is far to narrow
to describe all measurements procedures in quantum mechanics. This has led
to the formulation of generalized measurements or positive operator values
measures (POVMs). Mathematically, a POVM consists of a collection of effects
$\EE=\{E(i), i \in I\}$ which are positive, $E(i) \geq 0$, and sum up to the
identity, $\sum_i E(i) = \openone.$  The effects describe the measurement 
outcomes and the probability of an outcome $i$ is given by $p(i) = \tr[\vr E(i)].$
Physically, any POVM can be realized by first letting the physical system interact 
with an auxiliary system and then measuring an observable on the auxiliary system. 
Finally, any observable $A$ is also a POVM if one identifies the $E(i)$ with the 
projectors onto the eigenvectors of $A$, in which case the measurement is also called a
projection valued measure (PVM). 

Given the notion of generalized measurements the question arises, when 
two or more POVMs can be considered to be nonclassical. One possibility is
to require the commutativity of all the effects, but more refined notions 
are useful. Indeed, several notions such as ``non-disturbance'',
``joint measurability'', and ``coexistence'' have been introduced and 
they are an active area of research \cite{ReReWo, HeWo, Lahti02, Seminal1, Pellonpaa}.

In this paper we argue that the notion of joint measurability is the proper
one to describe classical behavior of two or more generalized measurements. 
To do so, we establish a connection between joint measurability and the task
of quantum steering. Quantum steering refers to the scenario, where one party, 
usually called Alice, wishes to convince the other party, called Bob, that
she can steer the state at Bob's side by making measurements on her side. 
This task was introduced by E. Schr\"odinger to demonstrate the puzzling
effects of quantum correlations \cite{erwin}  and recently, it has attracted an 
increasing attention again \cite{WiJoDo, pusey, SkNaCa, moroder14, bowles14a, piani14a}.

More precisely, we show that a set of POVMs is not jointly measurable if and
only if the set can be used for Alice to show the steerability of some quantum
state. This allows to derive new steering inequalities from results known 
for joint measurability, and we will also find new criteria for joint 
measurability from results on steering. Finally, we demonstrate that 
other possible extensions of commutativity to generalized measurements, 
such as coexistence, lead to nonclassical effects and we explore the 
relation of joint measurability to Bell inequality violations.

{\it Joint measurability ---} 
The notion of joint measurability is most conveniently introduced with
an example. The Pauli spin matrices $\sigma_x$ and $\sigma_z$ are noncommuting
and cannot be measured jointly. However, one can consider the smeared or
unsharp measurements $S_x$ and $S_z$, defined by the POVM elements 
$S_x(\pm)=\frac{1}{2}(\openone \pm\frac{1}{\sqrt2}\sigma_x)$ and 
$S_z(\pm)=\frac{1}{2}(\openone \pm\frac{1}{\sqrt2}\sigma_z)$. 
It was shown in Ref.~\cite{Paul86} that these are jointly measurable: 
one can consider the joint observable 
\begin{eqnarray}
G(i,j)=\frac{1}{4}(\openone+\frac{i}{\sqrt2}\sigma_x+\frac{j}{\sqrt2}\sigma_z),\ i,j\in\{-1,+1\}.
\end{eqnarray}
and since the $S_x(\pm)=\sum_j G(\pm, j)$ and $S_z(\pm)=\sum_i G(i,\pm)$
one can jointly determine the probabilities of the generalized measurements $S_x$ and
$S_z$ by measuring $G$. More generally, a set of POVMs $\{\EE_k\}$ 
is said to be jointly measurable iff there exists a POVM $G$ from which the probabilities of the POVM elements of the $\EE_k$ can be computed via classical
post-processing. This is, of course, the case if
all the effects of all POVMs commute, but the example from above shows that
joint measurability is more general than just the commutativity of the effects.

For our purposes it is important that joint measurability of the set
$\{E_k\}$ can be formulated as the existence of a set of positive 
operators $\{G(\lambda)\}$ from which the original observables can be 
attained as
\begin{eqnarray}
\sum_{\lambda} D_\lambda(x|k) G(\lambda) 
&=&E_k(x)\ \mbox{ for all } x,k,
\label{3.7.2014 12:44}
\end{eqnarray}
with $\sum_{\lambda}G(\lambda) = \openone$ and where 
$D_\lambda(x|k)$ are positive constants with 
$\sum_x D_\lambda(x|k)=1$ \cite{TwCaHeTo}. In practice, this means
that the probabilities of the results $E_k(x)$ can be determined by 
measuring the $G(\lambda)$ and classically post-processing the 
data.

{\it Quantum steering ---} The essence of steering can be described 
by a similar example. Let us assume that two parties, Alice and Bob, 
share a maximally entangled two-qubit state 
$\ket{\psi}=(\ket{00}+\ket{11})/\sqrt{2}.$ If Alice measures the Pauli operators $\sigma_x$ or $\sigma_z$, the state on Bob's side will be an eigenstate $\ket{x^\pm}$
or $\ket{z^\pm}$ depending on Alice's measurement and result. Since all these
states are pure, Bob cannot explain this by assuming that he has a fixed 
marginal state $\vr_B$ which is only modified due to the additional knowledge
from Alice's measurements. So Bob must conclude that Alice can steer the state
in his lab by making measurements on her side. The question arises whether
the same phenomenon occurs if Alice uses the smeared measurements $S_x$
and $S_z$ introduced above. This will be answered in full generality
in the following. 

First, we label Alice's and Bob's POVMs by $\{A_k\}$ and 
$\{B_l\}$ and the system's state by $\varrho_{AB}$. Clearly,  
the scenario is non-steerable if the probabilities of possible events 
can be written in the form
\begin{eqnarray}\label{22.6.2014 15:40}
\text{tr}[\varrho_{AB}A_k(x)\otimes B_l(y)]=\sum_{\lambda}p(\lambda)p(x|k,\lambda)\text{tr}[\varrho_\lambda B_l(y)]
\end{eqnarray}
because then Bob can assume that he has the collection of states 
$\varrho_\lambda$ with probabilities $p(\lambda)$ which is only modified
by additional information $p(x|a,\lambda)$ from Alice's measurement.
We can write the left hand side of this equation as
\begin{eqnarray}
\text{tr}[\text{tr}_A[(A_k(x)\otimes \openone)\varrho_{AB}] B_l(y)]
=:\text{tr}[\varrho_{x|k} B_l(y)]
\end{eqnarray}
and if the measurements of Bob are tomographically complete
it follows that $\varrho_{x|k}=\sum_{\lambda}p(\lambda)
p(x|a,\lambda)\varrho_\lambda$. If, on the other hand, 
the quantities $\varrho_{x|k}$ admit this kind of a 
decomposition (also called a hidden state model)
we conclude that the scenario is non-steerable.

This can be reformulated as suggested in Refs.~\cite{SkNaCa, pusey}:
Steering is equivalent to the non-existence of a set of positive 
operators $\{\sigma_\lambda\}$ such that
\begin{eqnarray}
\label{21.5.14 20:02}
\sum_\lambda p(x|k,\lambda)\sigma_\lambda&=&\varrho_{x|k}\ \mbox{ for all } x,k
\end{eqnarray}
with $\text{tr}(\sum_\lambda \sigma_\lambda)=1$  and 
where $\varrho_{x|k} =\text{tr}_A[(A_k(x)\otimes \openone)\varrho_{AB}]$ 
are Bob's not-normalized conditional states. The formal similarity
between Eq.~(\ref{3.7.2014 12:44}) and Eq.~(\ref{21.5.14 20:02}) is appealing 
and, as we will see now, no coincidence.

{\it Steering and joint measurements ---} Consider the case
where Alice has observables $\{A_k\}$ which are jointly 
measurable. Using Eq.~(\ref{3.7.2014 12:44}) we can write for 
any steering scenario the conditional states of Bob as
\begin{eqnarray}
\label{11.6.14 18:42}
\varrho_{x|k}
=\sum_{\lambda}D_\lambda(x|k)\text{tr}_A[(G(\lambda)\otimes \openone)\varrho_{AB}]
\end{eqnarray}
which is a decomposition as in Eq.~(\ref{21.5.14 20:02}). Therefore, if 
Alice's observables are jointly measurable then the scenario is 
non-steerable.

On the other hand, if the measurements are not jointly measurable, 
one can always find a state which can be used for steering: for 
the maximally entangled state 
$|\phi^+\rangle=\frac{1}{\sqrt d}\sum_{i=1}^d|ii\rangle$ 
one can write Bob's conditional states as
\begin{eqnarray}\label{21.5.14 18:05}
\varrho_{x|k}=\text{tr}_A[(A_k(x)\otimes \openone)
|\phi^+\rangle\langle\phi^+|]
=\frac{1}{d}[A_k(x)]^T.
\end{eqnarray}
If the scenario is not steerable then one can find a set of 
positive operators $\{\sigma_\lambda\}$ and a set 
of positive numbers $p(x|k,\lambda)$ such that
\begin{eqnarray}\label{21.5.14 18:20}
A_k(x)=d\sum_\lambda p(x|k,\lambda)
\sigma_\lambda^T=:\sum_\lambda D_\lambda(x|k)G(\lambda).
\end{eqnarray}
This is just the joint measurability condition from 
Eq.~(\ref{3.7.2014 12:44}). Note that the normalization condition also holds because the reduced state of $|\phi^+\rangle$ is maximally mixed.
So we can state:
\begin{theorem}\label{2.7.2014 14:48}
Generalized measurements are not jointly measurable if and only if
they can be used for quantum steering.
\end{theorem}
Let us note that the reasoning prior to Observation 1 was done for the maximally entangled state. Steering is however invariant under SLOCC on the characterized (Bob's) side. This means that any state which is obtained from the maximally entangled one by SLOCC can be used to show steering for not jointly measurable observables. Therefore any pure Schmidt rank $d$ state (possibly having an arbitrary small amount of entanglement) reveals steering.

We exploit the connection with two notes on different formulations of simultaneous measurability, give a generic incompatibility criteria for sharp observables, and show a steering inequality based on the Fermat-Torricelli point.

{\it Coexistence is nonclassical ---} Coexistence of POVMs $A_1$ and $A_2$ means the possibility of making a measurement $G$ of which statistics include the statistics of $A_1$ and $A_2$. To be more precise, $A_1$ and $A_2$ are coexistent if their POVM elements are contained in the range (i.e. all possible sums of POVM elements) of a third POVM $G$. Note that contrary to joint measurements the statistics do not need to originate from a post-processing scheme as in Eq.~(\ref{3.7.2014 12:44}). To clarify the notion we present an example given in Ref.~\cite{ReReWo} which was originally used to show that coexistence is more general than joint measurability; for a similar example Ref.~\cite{Pellonpaa}.

In $\mathbb C^3$ define $|\varphi\rangle=\frac{1}{\sqrt3}(|1\rangle+|2\rangle+|3\rangle)$ and a POVM $G$ by the elements $\{\frac{1}{2}|1\rangle\langle1|,\frac{1}{2}|2\rangle\langle2|,\frac{1}{2}|3\rangle\langle3|,\frac{1}{2}|\varphi\rangle\langle\varphi|,\frac{1}{2}(\openone -|\varphi\rangle\langle\varphi|)\}$. One sees straightforwardly that the measurement statistics of a 3-valued POVM $A_1$ defined as $A_1(i)=\frac{1}{2}(\openone -|i\rangle\langle i|)$ and a 2-valued POVM $A_2$ defined as $A_2(1)=\frac{1}{2}|\varphi\rangle\langle\varphi|,\ A_2(2)=\openone -A_2(1)$ are contained in the measurement statistics of $G$, hence they are coexistent. In Ref.~\cite{ReReWo} it was shown that these measurements are nevertheless not jointly measurable due to the lack of a post-processing relation. By Observation 1 we conclude the following:
\begin{theorem}
Coexistence of generalized measurements has a nonclassical feature, {\emph i.e.} it can reveal steering.
\end{theorem}

{\it Non-disturbance can be classical ---}
One way to define classicality of two measurements, say $A_1$ and $A_2$, is to say that the measurement of $A_1$ does not disturb the measurement of $A_2$. This means that a measurement of $A_1$ updates the state in such a way that a subsequent measurement of $A_2$ has the same statistics for both the updated and the original state. It was shown in Ref.~\cite{HeWo} that there exists pairs of observables that can be measured jointly even though they do not admit a non-disturbing sequential measurement. Using this together with Observation 1 we conclude that this nonclassical feature of disturbing measurement does not necessary lead to steering.

{\it From steering to incompatibility --} 
We show that there exists a threshold value of white noise that one needs to add in order to get any set of PVMs jointly measurable. For this purpose we need the following connection between noisy states and noisy observables:
\begin{eqnarray}
\text{tr}_A[A_k(x)\otimes\openone\varrho_{AB}^\lambda]=\text{tr}_A[A_k^\lambda(x)\otimes\openone\varrho_{AB}],
\label{25.6.2014 16:06}
\end{eqnarray}
where 
\begin{eqnarray}
\varrho_{AB}^\lambda&=&\lambda\varrho_{AB}+\frac{1-\lambda}{d}\openone\otimes\text{tr}_A[\varrho_{AB}],\\ 
A_k^\lambda(x)&=&\lambda A_k(x)+\frac{1-\lambda}{d}\text{tr}[A_k(x)]\openone.\label{2.7.2014 14:37}
\end{eqnarray}

In order to obtain the threshold value we take the known result from Ref.~\cite{WiJoDo} stating that the maximally entangled state is steerable with PVMs up to the amount $\lambda^*:=\frac{H_d-1}{d-1}$ of white noise, where $H_d=\sum_{n=1}^d\frac{1}{n}$. Using Eq.~(\ref{25.6.2014 16:06}) and Observation 1 on obtains that for any smearing parameter $\lambda\geq\lambda^*$ there must exist a set of PVMs which is noise resistant up to $\lambda$. On the other hand, the maximally entangled state reveals steering for not jointly measurable observables, so all PVMs must be jointly measurable with the amount $\lambda^*$ of white noise. Thus we arrive at the following result:
\begin{theorem}
In a $d$-dimensional Hilbert space any set of sharp observables is jointly measurable with the amount $\lambda^*$ of white noise. Moreover, for any amount of smearing above this limit there exists a set of PVMs which remains not jointly measurable.
\end{theorem}
Note that this is formerly known to be sufficient for $d=2$ \cite{MLQT}. The result leads to an interesting open question: do POVMs resist more noise than PVMs? If this is the case then PVMs are not enough for concluding steerability of a state and if it is not the case then this directly leads to new local hidden variable models for POVMs.

{\it Fermat-Torricelli steering inequality --} There are many results of joint measurability known in terms of white noise resistance \cite{Paul86, Spekkens, TeikoMUB}. As an example consider that Alice has three dichotomic unbiased (i.e. $p(\pm|k)=\frac{1}{2}$) measurements while Bob's conditional (normalized) qubit states are characterized by the Bloch vector $\vec x_k,\ k=1,2,3$. Using the joint measurability criterion of Ref.~\cite{YuOh} the observed data is steerable iff
\begin{eqnarray}\label{28.6.2014 13:03}
&&\|\vec x_1+\vec x_2+\vec x_3-\vec x_{FT}\|+\|\vec x_1-\vec x_2-\vec x_3-\vec x_{FT}\|\\
&&+\|\vec x_1-\vec x_2+\vec x_3+\vec x_{FT}\|+\|\vec x_1+\vec x_2-\vec x_3+\vec x_{FT}\|>4\nonumber,
\end{eqnarray}
where $\vec x_{FT}$ denotes the Fermat-Torricelli point of the vectors $\vec x_1+\vec x_2+\vec x_3,\vec x_1-\vec x_2-\vec x_3,-\vec x_1+\vec x_2-\vec x_3,$ and $-\vec x_1-\vec x_2+\vec x_3$, i.e. it is the vector that minimizes the sum in Eq.~(\ref{28.6.2014 13:03}).

\textit{Joint measurability and nonlocality ---}
{From} the previous discussion we know that any non 
jointly measurable set of POVMs can reveal its ``quantumness'' 
in a strictly nonclassical, nonlocal effect, more precisely, in 
the form of steering. Steering is however not the ultimate strongest form of nonlocality since one still needs a quantum description on one side. Thus it is of course a natural question whether this connection can even be strenghtened, so whether it also holds that any not jointly measurable set of POVMs can show nonclassicality in a Bell type scenario. 

This is indeed the case for two dichotomic measurements as has been shown by Wolf~\textit{et al.}~in Ref.~\cite{WoPeFe}. It also holds for an arbitrary number of PVMs. In the following, we argue that this connection might be very surprising to hold in generality, since via a very simple example one encounters already large difficulties.

Consider the three dichotomic spin measurements of a qubit $A^{\lambda}_{k}(\pm)=(\openone\pm\lambda \sigma_k)/2$ with $k \in \{x,y,z\}$. As already mentioned the additional parameter $\lambda$ characterizes the noise on these measurements. For $\lambda=1$ the measurements  $A_k:= A_k^{\lambda=1}$ are noncommmuting projectors, while for $\lambda \leq 1/\sqrt{3}\approx 0.5774$ the set of POVMs becomes jointly measurable. Suppose that joint measurability and nonlocality are as strongly connected as steering. This would mean that for any noisy, but not jointly measurable set of these POVMs, \textit{i.e.}, for all $1/\sqrt{3} < \lambda $, it is possible to find a respective bipartite state $\varrho_{AB}$ and corresponding measurements for Bob $B_l(k)$, such that the obtained data $P(\pm,y|k,l)=\TR[\varrho_{AB}A_k^{\lambda}(\pm) \otimes B_l(y)]$ violate a Bell inequality. 

In the search for such an appropriate state, first note that pure states $\varrho_{AB}=\ket{\psi}\bra{\psi}$ are sufficient, since any mixed state can only violate a Bell inequality if at least one pure state out of its range does so. Using the Schmidt decomposition and because  $\dim(\mathcal{H}_A)=2$ the most general pure state is given by $\ket{\psi}=U_A \otimes U_B \ket{\psi_s}$ with $\ket{\psi_s}=s\ket{00}+\sqrt{1-s^2}\ket{11}$ where $1/\sqrt{2}\leq s \leq 1$. Since we optimize Bob's measurements we can  additionally assume $U_B=\openone$, meaning that Bob similarly holds a qubit. Next, we also like to transfer the noise of the measurements into the state, as given by Eq.~(\ref{25.6.2014 16:06}).
Thus, rather looking for a pure state which violates a Bell inequality using the noisy measurements $A^{\lambda}_k$, we can equivalently search for a mixed state that violates a Bell inequality with perfect measurements $A_k$. To sum up, we would need to show that for any parameter $\lambda > 1/\sqrt{3}$, a state of the form
\begin{eqnarray}
\nonumber
\varrho_{AB}(s;U_A) &=& \lambda U_A \otimes \openone\ket{\psi_s}\bra{\psi_s} U_A^\dag \otimes \openone \\
\label{eq:rhoS}
&&+ (1-\lambda) \openone/2 \otimes \TR_{A}[\ket{\psi_s}\bra{\psi_s}]
\end{eqnarray}
with appropriately chosen $1/\sqrt{2}\leq s \leq 1$ and $U_A$ violates a Bell inequality using the three perfect spin measurements on system $A$, and arbitrary measurements for system $B$. 

Let us start with the maximally entangled state, $s=1/\sqrt{2}$, for which it is known that it does not violate a Bell inequality using projective measurements if $\lambda \leq 0.6595$~\cite{grothendiek}. Hence, for the given noisy not jointly measurable set of POVMs within $1/\sqrt{3} < \lambda \leq 0.6595$, the data of the maximally entangled state, using also projective measurements for Bob, will not display any nonlocality. For non-maximally entangled states the situation is much less analyzed, especially under the influence of coloured noise as in Eq.~(\ref{eq:rhoS}). The statement extends however to $1/\sqrt{3} < \lambda \leq 0.6009$~\cite{grothendiek} for arbitrary, non-maximally entangled states if one wants to reproduce the full correlations. Thus the only Bell inequalities that remain are the ones with marginals.

\begin{figure}[t]
  \begin{center}
    \includegraphics[angle=-90,scale=0.30]{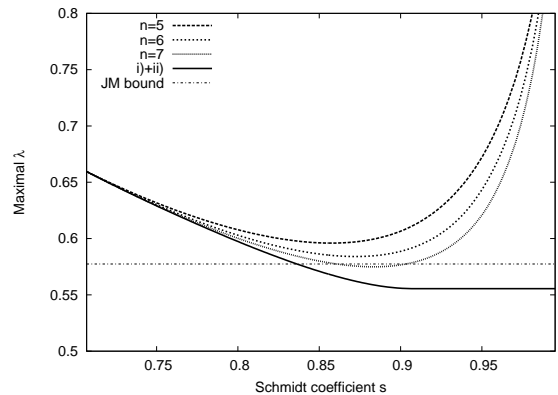}
  \end{center}
  \caption{Maximal values of $\lambda$ when a decomposition as given by Eq.~(\ref{eq:strategy_LHV}) is possible for all $U_A$ depending on the Schmidt coefficient $s$. It shows that a pure state with $s\leq 0.835$ is never able to reveal Bell nonlocality for an arbitrary number of projective measurements, while for $n\leq6$ projective measurements it is not possible for any state.}
  \label{fig:noise_parameters}
\end{figure}

A different way to prove that certain states do not violate a Bell inequality is to write them as a convex combination of states known to possess a local hidden variable model for the considered configuration, 
\begin{equation}
\label{eq:strategy_LHV}
\varrho_{AB}(s;U_A) = \sum_i p_i \varrho_i^{\rm LHV}.
\end{equation}
Generic states that we consider in this decomposition include: (i) noisy Bell states with $\lambda\leq 0.6595$ and (ii) states with $2$ symmetric extensions of system $A$. States of the class (ii) are known to have a local hidden variable model for three generic measurements for system~$A$~\cite{terhal}, such that we exploit the fact that Alice has only a restricted number of measurements. Such a search can be easily done with semidefinite programming~\cite{sdp}. Fig.~\ref{fig:noise_parameters} shows, depending on the Schmidt coefficient $s$ (and for all $U_A$), the respective maximal values of $\lambda$ when such a decomposition is possible. As can be seen for $s \leq 0.835$, there is always a noise parameter $\lambda > 1/\sqrt{3}$ such that the given set of POVMs is not jointly measurable, but the measured state will not violate a Bell inequality using an arbitrary number of projective measurements for Bob. At last, if one additionally constrains Bob to perform only $n$ different dichotomic measurements then one can further add 
(iii) the class of states that have $n-1$ symmetric extensions for system $B$. As shown in Fig.~\ref{fig:noise_parameters} for $n \leq6$ such a decomposition is possible for all values of $s$. Thus, there exists a parameter $\lambda>1/\sqrt{3}$ such that the corresponding set of POVMs is not jointly measurable but no quantum state will display nonlocality if Bob only carries out $6$ dichotomic measurements. 

We are confident that these observations give strong hints that there are sets of POVMs which are not jointly measurable, but which are nevertheless useless to certify nonlocality. 

\textit{Conclusions ---} 
We have shown that joint measurability and quantum steering are intrinsically connected: A collection of different measurements are not jointly measurable if and only if they can reveal its ``nonclassicality'' as a violation of a steering inequality. This connects the abstract notion of joint measurability to an explicit nonlocality task, and thereby singles out not joint measurability as a special nonclassical property among other peculiar quantum features of measurements. 

Since measurements are as central as quantum states, we believe that this connection will spur the resource theory of measurements, \textit{i.e.}, which kind of measurements are required for certain tasks. This investigation could provide some operational meaning to other quantum properties of measurements such as disturbance or non-coexistence. 

\begin{acknowledgments}
We would like to thank C.~Budroni, T.~Heinosaari, M.~Huber, and J.~Zuniga for their support. 
This work has been supported by the EU (Marie Curie CIG 293993/ENFOQI), the BMBF (Chist-Era Project QUASAR), 
the FQXi Fund (Silicon Valley Community Foundation) and the DFG. RU acknowledges Finnish Cultural Foundation for financial support.
\end{acknowledgments}

\textit{Note added.}---After finishing this work we noticed that similar results were obtained in Ref.~\cite{quintino14a}.

\end{document}